\documentclass[twocolumn]{aastex62}

\usepackage{amsmath}
\usepackage{bold-extra}
\usepackage{hyperref}
\usepackage{xcolor}
\usepackage{float}
\usepackage{placeins}
\usepackage{gensymb}

\newcommand{\othree}{[O~{\sc iii}]}
\newcommand{\mgtwo}{Mg~{\sc ii}}
\newcommand{\cfour}{C~{\sc iv}}
\newcommand{\hb}{H$\beta$}
\newcommand{\ha}{H$\alpha$}
\newcommand{\fetwo}{Fe~{\sc ii}}
\newcommand{\fethree}{Fe~{\sc iii}}
\newcommand{\hd}{H$\delta$}
\newcommand{\hg}{H$\gamma$}
\newcommand{\hetwo}{He~{\sc ii}}
\newcommand{\otwo}{[O~{\sc ii}]}
\newcommand{\nethree}{[Ne~{\sc iii}]}
\newcommand{\ntwo}{[N~{\sc ii}]}

\begin{document}

\title{Placing High-Redshift Quasars in Perspective: a Catalog of Spectroscopic Properties from the Gemini Near Infrared Spectrograph - Distant Quasar Survey}

\email{brandonmatthews@my.unt.edu}

\author{Brandon M. Matthews}
\affil{Department of Physics, University of North Texas,
Denton, TX 76203, USA}

\author{Ohad Shemmer}
\affil{Department of Physics, University of North Texas,
Denton, TX 76203, USA}

\author{Cooper Dix}
\affil{Department of Physics, University of North Texas,
Denton, TX 76203, USA}

\author{Michael S. Brotherton}
\affil{Department of Physics and Astronomy, University of Wyoming,
Laramie, WY 82071, USA}

\author{Adam D. Myers}
\affil{Department of Physics and Astronomy, University of Wyoming,
Laramie, WY 82071, USA}

\author{I. Andruchow}
\affil{Facultad de Ciencias Astron\'{o}micas y Geofi\'{i}sicas, Universidad Nacional de La Plata, Paseo del Bosque, B1900FWA La Plata, Argentina}
\affil{Instituto de Astrof\'{i}sica de La Plata, CONICET–UNLP, CCT La Plata, Paseo del Bosque, B1900FWA La Plata, Argentina}

\author{W. N. Brandt}
\affil{Department of Astronomy and Astrophysics, The Pennsylvania State University, University Park, PA 16802, USA}
\affiliation{Institute for Gravitation and the Cosmos, The Pennsylvania State University, University Park, PA 16802, USA}
\affiliation{Department of Physics, 104 Davey Lab, The Pennsylvania State University, University Park, PA 16802, USA}

\author{Gabriel A. Ferrero}
\affil{Facultad de Ciencias Astron\'{o}micas y Geofi\'{i}sicas, Universidad Nacional de La Plata, Paseo del Bosque, B1900FWA La Plata, Argentina}
\affil{Instituto de Astrof\'{i}sica de La Plata, CONICET–UNLP, CCT La Plata, Paseo del Bosque, B1900FWA La Plata, Argentina}

\author{S. C. Gallagher}
\affil{Department of Physics \& Astronomy, University of Western Ontario, 1151 Richmond St, London, ON N6C 1T7, Canada}

\author{Richard Green}
\affil{Steward Observatory, University of Arizona, 933 N Cherry Ave, Tucson, AZ 85721, USA}

\author{Paulina Lira}
\affil{Departamento de Astronom\'{i}a, Universidad de Chile, Casilla 36D, Santiago, Chile}

\author{Richard M. Plotkin}
\affil{Department of Physics, University of Nevada, Reno, NV 89557, USA}

\author{Gordon T. Richards}
\affil{Department of Physics, Drexel University, 32 S. 32nd Street, Philadelphia, PA 19104, USA}

\author{Jessie C. Runnoe}
\affil{Department of Physics \& Astronomy, Vanderbilt University, 6301 Stevenson Center Ln, Nashville, TN 37235, USA}

\author{Donald P. Schneider}
\affil{Department of Astronomy and Astrophysics, The Pennsylvania State University, University Park, PA 16802, USA}
\affiliation{Institute for Gravitation and the Cosmos, The Pennsylvania State University, University Park, PA 16802, USA}

\author{Yue Shen}
\altaffiliation{Alfred P. Sloan Research Fellow}
\affiliation{Department of Astronomy, University of Illinois at Urbana-Champaign, Urbana, IL 61801, USA}
\affiliation{National Center for Supercomputing Applications, University of Illinois at Urbana-Champaign, Urbana, IL 61801, USA}

\author{Michael A. Strauss}
\affil{Department of Astrophysical Sciences, Princeton University, Princeton, NJ 08544, USA}

\author{Beverley J. Wills}
\affil{University of Texas Astronomy Department, University of Texas at Austin, C1400, 1 University Station, Austin, TX 78712, USA}

\received{2020 August 12}
\accepted{2020 October 19}

\begin{abstract}

We present spectroscopic measurements for 226 sources from the Gemini Near Infrared Spectrograph - Distant Quasar Survey (GNIRS-DQS). Being the largest uniform, homogeneous survey of its kind, it represents a flux-limited sample ($m_{i}\lesssim19.0$ mag, $H\lesssim16.5$ mag) of Sloan Digital Sky Survey (SDSS) quasars at \hbox{$1.5 \lesssim z \lesssim 3.5$} with a monochromatic luminosity \hbox{($\lambda L_{\lambda}$)} at ~\hbox{5100 \AA} in the range of $10^{44} - 10^{46}$\,erg\,s$^{-1}$. A combination of the GNIRS and SDSS spectra covers principal quasar diagnostic features, chiefly the \cfour ~$\lambda1549$, \mgtwo ~$\lambda\lambda2798, 2803$, \hb ~$\lambda4861$, and \othree ~$\lambda\lambda4959, 5007$ emission lines, in each source. The spectral inventory will be utilized primarily to develop prescriptions for obtaining more accurate and precise redshifts, black hole masses, and accretion rates for all quasars. Additionally, the measurements will facilitate an understanding of the dependence of rest-frame ultraviolet-optical spectral properties of quasars on redshift, luminosity, and Eddington ratio, and test whether the physical properties of the quasar central engine evolve over cosmic time.

\end{abstract}


\section{Introduction} \label{sec:intro}

A persistent problem in extragalactic astrophysics is understanding how supermassive black holes (SMBHs) and their host galaxies co-evolve over cosmic time \citep[e.g.,][]{2008ApJ...676...33D,2010ApJ...708..137M,2011ARA&A..49..373B,2014ARA&A..52..589H}. This problem touches upon several aspects of galaxy evolution, including the SMBH mass ($M_{\rm BH}$), which correlates with properties of the host galaxy, such as the bulge mass and stellar velocity dispersion \citep[e.g.,][]{2000ApJ...539L...9F,2000ApJ...539L..13G,2010ApJ...716..269W,2013ARA&A..51..511K,2013ApJ...764..184M,2015ApJ...813...82R}, the accretion rate, which probes the accretion flow and efficiency of the accretion process, \citep[e.g.,][]{2006MNRAS.365...11C,2010MNRAS.407.1529H,2014SSRv..183...21B}, and the kinematics of material outflowing from the vicinity of the SMBH, which may affect star formation in the host galaxy \citep[e.g.,][]{2010MNRAS.401....7H,2012MNRAS.425L..66M,2018agn..confE..68C}. For nearby ($z \lesssim 1$) active galactic nuclei (AGNs) or quasars, most of the parameters required for exploring these topics can be most reliably estimated using optical diagnostics, namely the broad \hb ~$\lambda$4861 and narrow \othree ~$\lambda\lambda$4959, 5007 emission lines. However, at $z \gtrsim 1$, which includes the epoch of peak quasar activity (from $z = 1-3$), these diagnostic emission lines are redshifted beyond $\lambda_{\rm obs} \sim1\mu$m, firmly into the near-infrared (NIR) regime. Since the vast majority of large spectroscopic quasar surveys have been limited to $\lambda_{\rm obs} \lesssim1\mu$m, investigations of  large samples of quasars at $z \gtrsim 1$ are usually forced to use spectroscopic proxies for \hbox{\hb ~and \othree.} Using indirect proxies can lead not only to inaccurate redshifts \citep[e.g.,][]{1982ApJ...263...79G,2010MNRAS.405.2302H,2016ApJ...833...33D,2016ApJ...831....7S,2020ApJ...893...14D}, but also to systematically biased and imprecise estimates of fundamental parameters such as $M_{\rm BH}$ and accretion rate \citep[e.g.,][]{2012MNRAS.427.3081T,2012ApJ...753..125S,2016ApJS..224...14D}.

NIR spectra have been obtained for a few hundred quasars at $z \gtrsim 1$, but these spectra constitute a heterogeneous collection of relatively small samples \hbox{($\approx 10 - 100$ sources)} that span wide ranges of source-selection criteria, instrument properties, spectral band and resolution, and signal-to-noise ratio ($S/N$) \citep[e.g.,][]{1999ApJ...514...40M,2004ApJ...614..547S,2004A&A...423..121S,2007ApJ...671.1256N,2011ApJ...730....7T,2012ApJ...753..125S,2015ApJ...799..189Z,2016A&A...594A..91L,2016MNRAS.460..187M,2016ApJ...817...55S,2017MNRAS.465.2120C}. Thus, the current NIR spectroscopic inventory for high-redshift quasars is biased in a multitude of selection criteria, and none of these mini-surveys are capable of providing a coherent picture of SMBH growth across cosmic time.

To mitigate the various systematic biases present in the current NIR spectroscopic inventory, we have obtained NIR spectra of 272 quasars at high redshift using the Gemini Near-Infrared Spectrograph (GNIRS, \citeauthor{2006SPIE.6269E..4CE} \citeyear{2006SPIE.6269E..4CE}), at the Gemini-North Observatory, with a Gemini Large and Long Program\footnote{\url{http://www.gemini.edu/node/12726}}. By utilizing spectroscopy in the $\sim$ 0.8–2.5 $\mu$m band of a uniform, flux-limited sample of optically selected quasars at \hbox{$1.5 \lesssim z \lesssim 3.5$}, our Distant Quasar Survey (GNIRS-DQS) was designed to produce spectra that, at a minimum, encompass the essential \hb ~and \othree~region in each source while having sufficient $S/N$ in the NIR band to obtain meaningful measurements of this region. This survey assembles the largest uniform sample of $z \gtrsim 1$ quasars with rest-frame optical spectroscopic coverage. The spectral inventory presented in this catalog will allow development of single-epoch prescriptions, as opposed to \cfour ~reverberation mapping, for rest-frame ultraviolet (UV) analogs of key properties such as $M_{\rm BH}$ and accretion rate, along with revised redshifts based primarily on emission lines in the rest-frame optical band.

This paper describes the GNIRS observations and structure of the catalog; subsequent investigations will present the scientific analyses enabled by this catalog. Section~\ref{sec:targets} describes the target selection, and Section~\ref{sec:observations} describes the GNIRS observations, and the spectroscopic data processing. Section~\ref{sec:spec} presents the catalog of basic spectral properties, along with a smaller catalog of additional features that can be measured reliably in some of the spectra. Section~\ref{sec:app} summarizes the main properties of our catalog as well as comments on its future applications. Throughout this paper we adopt a flat $\Lambda$CDM cosmology with $\Omega_\Lambda = 1 - \Omega_0 = 0.7$ and $H_0 = 70$ ${\rm km s^{-1}} {\rm Mpc^{-1}}$ \citep{2007ApJS..170..377S}.

\section{Target Selection} \label{sec:targets}

\begin{figure}[H]
\includegraphics[scale=0.5]{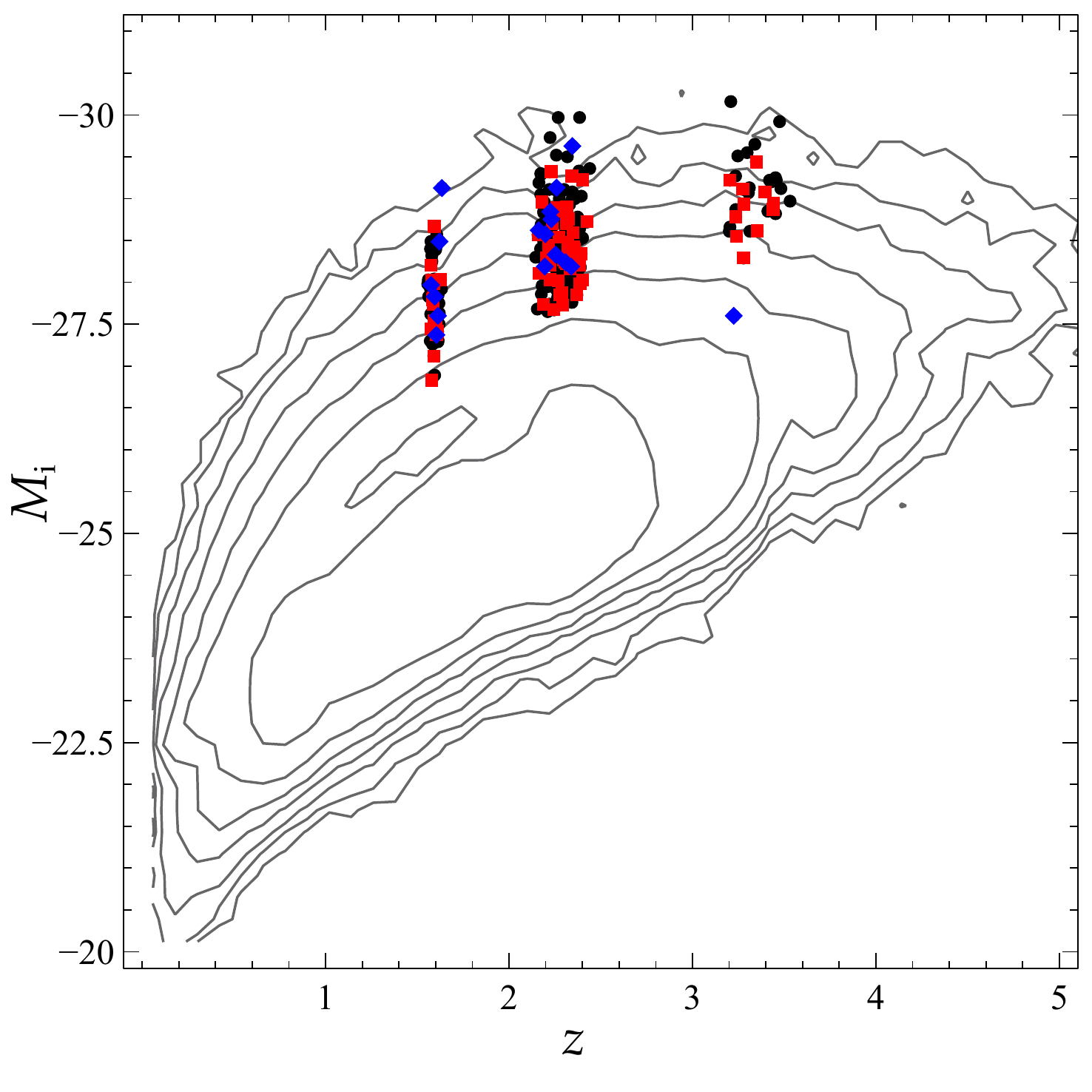}
\caption{Distribution of SDSS quasars from DR14 (contours) and the 272 objects in the GNIRS-DQS sample (symbols) in the luminosity-redshift plane, where $M_i$ is the absolute $i$-band luminosity (BAL quasars are represented by red squares, and non-radio quiet quasars are represented by blue diamonds). Most, but not all, quasars in DR14 are represented via contour lines, for clarity. Redshift ranges were chosen to ensure the prominent emission lines of \hb ~and \othree ~would be centered in the $J$, $H$, or $K$ band. The final sample is representative of the quasar population within our selection criteria.}
\label{fig:sdss}
\end{figure}

\begin{figure}[H]
\includegraphics[scale=0.3]{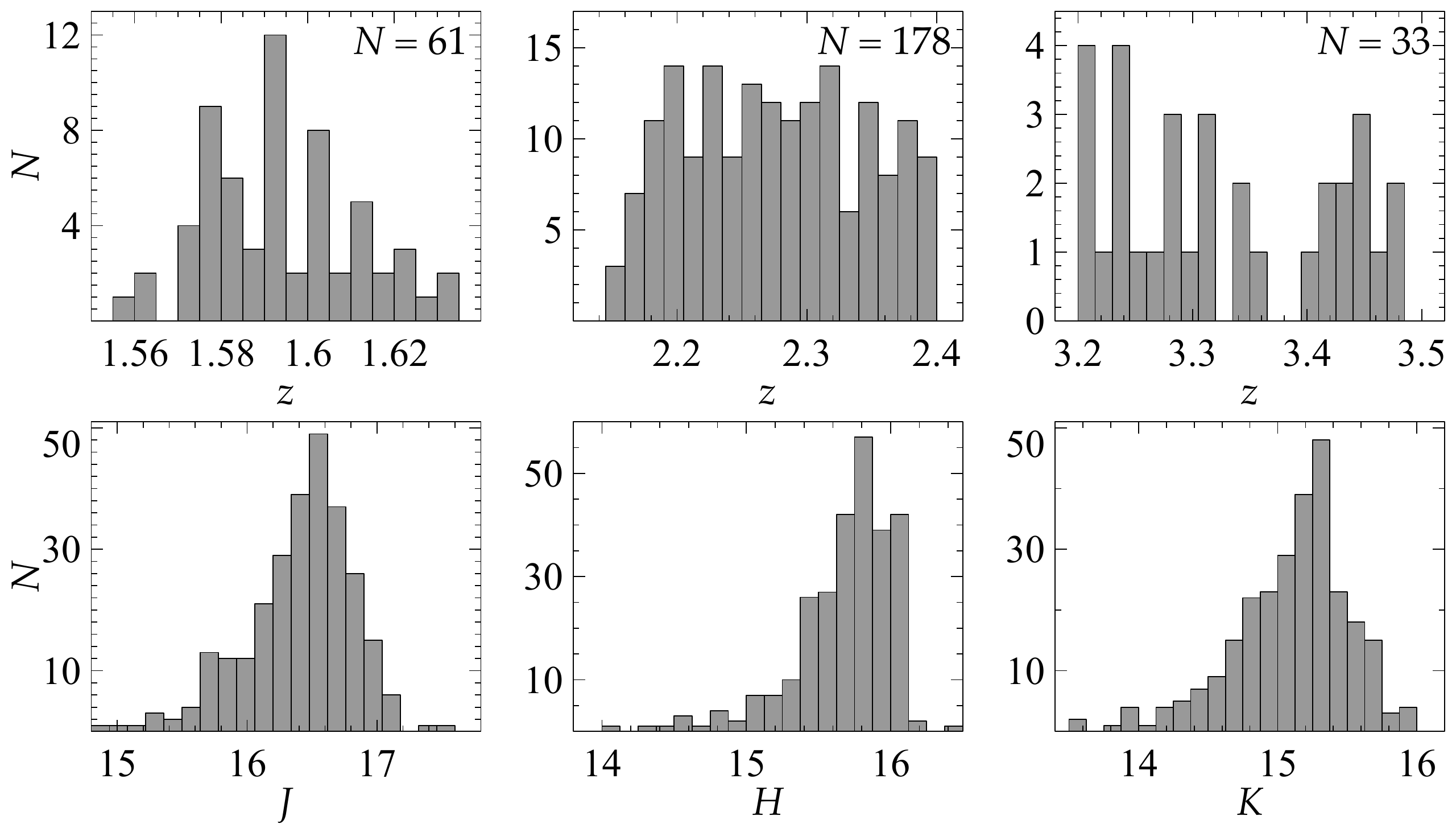}
\caption{Redshift distribution in each redshift interval from SDSS (top), and corresponding magnitude distribution of the 272 objects in our sample (bottom). The three redshift bins correspond to the \hb ~and \othree ~lines appearing at the center of the $J$, $H$, or $K$ photometric bands.}
\label{fig:jhk}
\end{figure}

The GNIRS-DQS targets were selected from the spectroscopic quasar catalog of the Sloan Digital Sky Survey \citep[SDSS;][]{2000AJ....120.1579Y}, primarily from SDSS Data Release 12 \citep[][]{2017A&A...597A..79P} and supplemented by SDSS Data Release 14 \citep[DR14;][]{2018A&A...613A..51P}. Sources were selected to lie in three narrow redshift intervals, \hbox{1.55 $\lesssim z \lesssim$ 1.65}, \hbox{2.10 $\lesssim z \lesssim$ 2.40}, and \hbox{3.20 $\lesssim z \lesssim$ 3.50}, in order to cover the \hbox{\hb+\othree} emission complex, and in order of decreasing NIR brightness, down to \hbox{$m_i$ $\sim$ 19.0}, a limit at which the SDSS is close to complete in each of those redshift intervals \citep{2002AJ....123.2945R}. Figure~\ref{fig:sdss} displays the luminosity-redshift distribution of GNIRS-DQS sources with respect to sources from the SDSS DR14 catalog. For the redshift distributions in the selected intervals, shown in Figure~\ref{fig:jhk} along with their respective magnitude distributions, the \hbox{\hb+\othree} emission complex reaches the highest $S/N$ in the centers of the $J$, $H$, and $K$ bands, respectively. The selected redshift intervals also ensure coverage of sufficient continuum emission and \fetwo~line emission flanking the \hbox{\hb+\othree} complex, enabling accurate fitting of these features. We visually inspected the SDSS spectrum of each candidate and removed sources having spurious redshifts, instrumental artifacts, and other anomalies. The combined SDSS-GNIRS spectroscopic coverage of each source includes, at a minimum, the \cfour~$\lambda$1549, \mgtwo ~$\lambda\lambda$2796, 2803, \hb, and \othree ~emission lines; the \ha ~$\lambda$6563 emission line is present in all sources at \hbox{1.55 $\lesssim z \lesssim$ 2.50}, representing $\sim$ 87\% of our sample. We note that the $2.10 \lesssim z\lesssim 2.40$ redshift bin comprises $\sim 67\%$ of our entire sample, given that this redshift bin is three times wider than that of the lower redshift bin, and sources in this bin are brighter than the sources in the higher redshift bin.

\begin{figure}[H]
\includegraphics[scale=0.5]{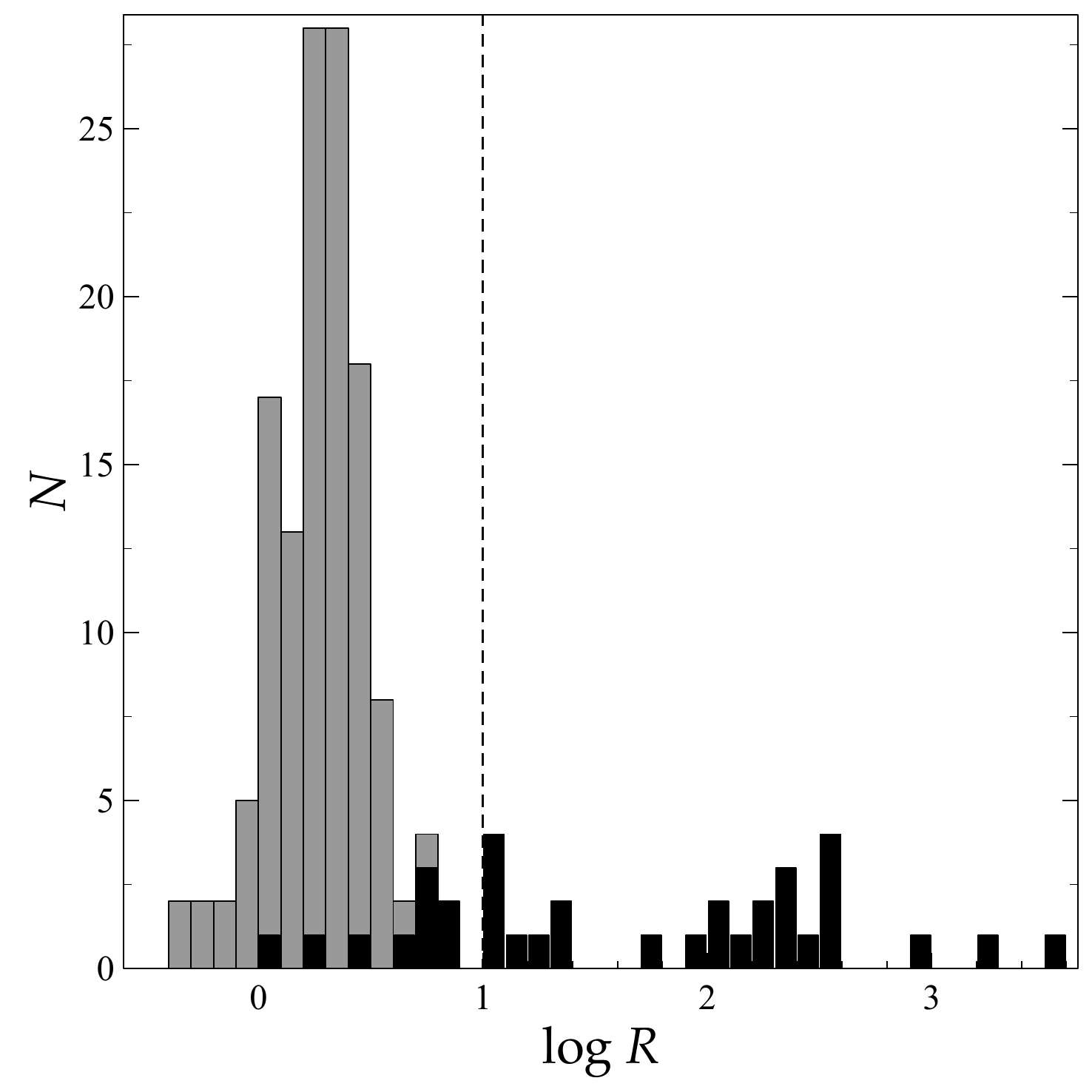}
\caption{Radio loudness distribution of the GNIRS-DQS sources; the shaded (grey) columns represent upper limits on $R$ for radio undetected sources based on the \cite{2018A&A...613A..51P} catalog, and the dashed line at $\log R=1$ incidates the threshold for radio quiet quasars. This distribution is generally similar to that of the SDSS quasar population \citep[e.g.,][]{2006AJ....131.2766R}.}
\label{fig:radio}
\end{figure}

In summary, the GNIRS-DQS sources constitute an optically-selected, NIR flux limited sample of quasars, spanning wide ranges in rest-frame UV spectral properties, including broad absorption line (BAL) and non-radio quiet quasars\footnote{\label{note:radio}We consider radio-quiet quasars to have $R <$ 10, where $R$ is the radio loudness, defined as $R=f_{\nu}$(5~GHz) / $f_{\nu}$ (4400~\AA), where $f_{\nu}$(5~GHz) and $f_{\nu}$(4400~\AA) are the flux densities at rest-frame frequencies of 5~GHz and 4400~\AA, respectively \citep{1989AJ.....98.1195K}. Non-radio quiet quasars include radio-intermediate ($10 < R < 100$) and radio-loud ($R > 100$) sources, respectively.}~\citep[comprising $\sim$ 30$\%$\footnote{Quasars flagged as BAL quasars in \cite{2018A&A...613A..51P} (see, Table~\ref{tab:obslog}).} and $\sim$ 12$\%$ of the sample, respectively;][]{2018A&A...613A..51P}. Figure~\ref{fig:radio} shows the radio loudness distribution of the GNIRS-DQS sources. The GNIRS-DQS sample is broadly representative of the general quasar population of luminous, high-redshift quasars during the epoch of most intense quasar activity \citep[e.g.,][]{1993ApJ...406L..43H,2005A&A...441..417H,2006AJ....131.2766R}.

\section{Observations, and Data Reduction} \label{sec:observations}

The observations were designed to yield data of roughly comparable quality, in terms of both $S/N$ and spectral resolution, to the respective SDSS spectra at \hbox{$\lambda_{\rm obs}\sim$~5000 \AA}. The GNIRS spectra were thus required to have a ratio of $\sim$ 40 between the mean flux density and the standard deviation of that flux density in a rest-frame wavelength interval spanning 100~\AA\ around $\lambda_{\rm rest} = 5100$\,\AA, and a spectral resolution of \hbox{$R \sim1100$} across the entire GNIRS band. These requirements enable accurate measurements of redshift based on \othree ~line peaks, with the high $S/N$ contributing to reducing the uncertainties below the spectral resolution limit, \hbox{$\sim$ 300 km s\textsuperscript{-1}} \citep{2016ApJ...831....7S}. As explained below in Section~\ref{sec:spec}, we determine that, on average, our spectra produce uncertainties on the measured line peak of \othree ~$\lambda 5007$ of order $\sim 50$ km s$^{-1}$, stemming from pixel-to-wavelength calibration and our fitting procedures.

All spectra were obtained in queue observing mode with GNIRS configured to use the Short Blue camera (0.15\arcsec pix\textsuperscript{-1}), the 32 lines mm\textsuperscript{-1} grating in cross-dispersed mode, and the 0.45\arcsec-wide slit. This configuration covers the observed-frame $\sim$0.8–2.5 $\mu$m band in each source, simultaneously, in six spectral orders with overlapping spectral coverage. Our observing strategy utilized an ABBA method of slit nodding to enable sky subtraction. Exposure times ranged from $\sim10-40$ minutes for each object, with an additional 15 minutes of overhead per source. Each observation included calibration exposures, and either one or two ABBA sequences depending on source brightness. We also observed a telluric standard star either immediately before or after the observation in a spectral range of B8\,V to A4\,V, with \hbox{8200\,K $\lesssim T_{\rm eff}\lesssim$ 13000\,K}, and typically within $\approx 10\degree-15\degree$ from each quasar.

The observation log of the original 272 sources appears in Table~\ref{tab:obslog}. Column (1) is the SDSS designation of the quasar. Column (2) provides the most reliable reported redshift estimate from SDSS \citep[][Table A1, column 9 ``Z"]{2018A&A...613A..51P}.  Columns (3), (4), and (5) list the respective $J$, $H$, and $K$ magnitudes of each quasar from the Two Micron All Sky Survey (2MASS; \citeauthor{2006AJ....131.1163S} \citeyear{2006AJ....131.1163S}). Columns (6) and (7) give the observation date and semester, respectively. Column (8) is the net science exposure time, Column (9) provides comments, if any, concerning the observation, Column (10) provides a flag for whether or not the quasar is a BAL quasar \citep[as defined in][]{2018A&A...613A..51P}, and Column (11) provides a flag for whether or not the quasar is considered non-radio quiet (see, footnote~\ref{note:radio}). 

We classify an acceptable observing night for this survey based on our programs' approved observing conditions including no greater than 50\% cloud cover and 85\% image quality\footnote{\url{https://www.gemini.edu/observing/telescopes-and-sites/sites\#Constraints}}, however some objects were observed under worse conditions, and are noted as such in Table~\ref{tab:obslog}. Additionally, 12 sources were observed over two observing sessions. These additional observations are recorded separately and immediately follow the initial observation in Table~\ref{tab:obslog} (which brings the total number of lines in that Table to 284). For these objects, all available observations were utilized in the reduction process.

Our data processing procedure generally follows the XDGNIRS pipeline developed by the Gemini Observatory (\citeauthor{2015ApJS..217...13M} \citeyear{2015ApJS..217...13M}\footnote{\url{http://www.gemini.edu/sciops/instruments/gnirs/data-format-and-reduction/reducing-xd-spectra}}: see also \citeauthor{2019ApJ...873...35S} \citeyear{2019ApJ...873...35S}) with the Gemini package in PyRAF\footnote{\url{https://www.gemini.edu/node/11823}}. Following standard image cleaning for artifacts and other observational anomalies, we pair-subtract the images to remove the bulk of the background noise by directly combining the sky-subtracted object exposures. Quartz lamps and IR lamps were used to create flat fields to correct pixel-by-pixel variation across the detector. The flat-fielded images were corrected for optical distortions. Several objects required replacement flat fields due to pixel shifting of dead pixels in the detector into the GNIRS spectra directly (marked accordingly with a corresponding comment in Table~\ref{tab:obslog}), which produced a notable increase in the uncertainty of spectroscopic measurements for these objects, particularly in the bluer bands. On average, the increased flux uncertainty from these spectra is on the order of $\sim$3\%. At this stage, of the 272 sources observed, 46 observations did not yield a meaningful spectrum due to bad weather, instrument artifacts, or other technical difficulties (Note 4 in Column 9 of Table~\ref{tab:obslog}), leaving the final sample at 226 sources.

Wavelength calibration was performed using two argon lamp exposures in order to assign wavelength values to the observed pixels. The uncertainties associated with this wavelength calibration are not larger than 0.5~\AA ~RMS, corresponding to \hbox{$\lesssim 10$ $\rm{km}$ $\rm{s^{-1}}$} at \hbox{$\sim15000$ \AA}.

Spectra of the telluric standards were processed in a similar fashion, followed by a careful removal of the stars' intrinsic hydrogen absorption lines. This process was performed by fitting Lorentzian profiles to the hydrogen absorption lines, and interpolating across these features to connect the continuum on each side of the line. Following the line cancellation, the quasar spectra were divided by the corrected stellar spectra. The corrected spectra were multiplied by an artificial blackbody curve with a temperature corresponding to the telluric standard star, which yielded a cleaned, observed-frame quasar spectrum. Each quasar spectrum was flux calibrated by comparing local flux densities to the $J$, $H$, and $K$ 2MASS magnitudes from Table~\ref{tab:obslog} and using the magnitude-to-flux conversion factors from Table A.2 of \citet{1998A&A...333..231B}. For the final spectra, we masked any noise present from cosmic rays, regions of high levels of atmospheric absorption, and band gap interference. 

We chose this method as opposed to flux calibrating via the telluric standards to avoid any differences in atmospheric conditions between observations of the object and the telluric standard. This preference was also motivated by our use of a relatively narrow slit in order to prioritize spectral resolution at the cost of potentially larger slit losses in the observations. Although the 2MASS and Gemini observations are separated by several years in the quasars' rest frames, the cross-calibrations are subject to minimal uncertainties since $\sim88\%$ of our sources are luminous radio-quiet quasars at high redshift. Such sources typically show UV-optical flux variations on the order of $\lesssim10$\% over such timescales \citep[e.g.,][]{2004ApJ...601..692V,2007ApJ...659..997K,2012ApJ...753..106M}. In fact, the effects stemming from the differences in airmass between the quasars and their respective telluric standard stars, as well as the slit losses, are typically larger than the expected intrinsic quasar variability.

\begin{figure}[H]
\includegraphics[scale=0.5]{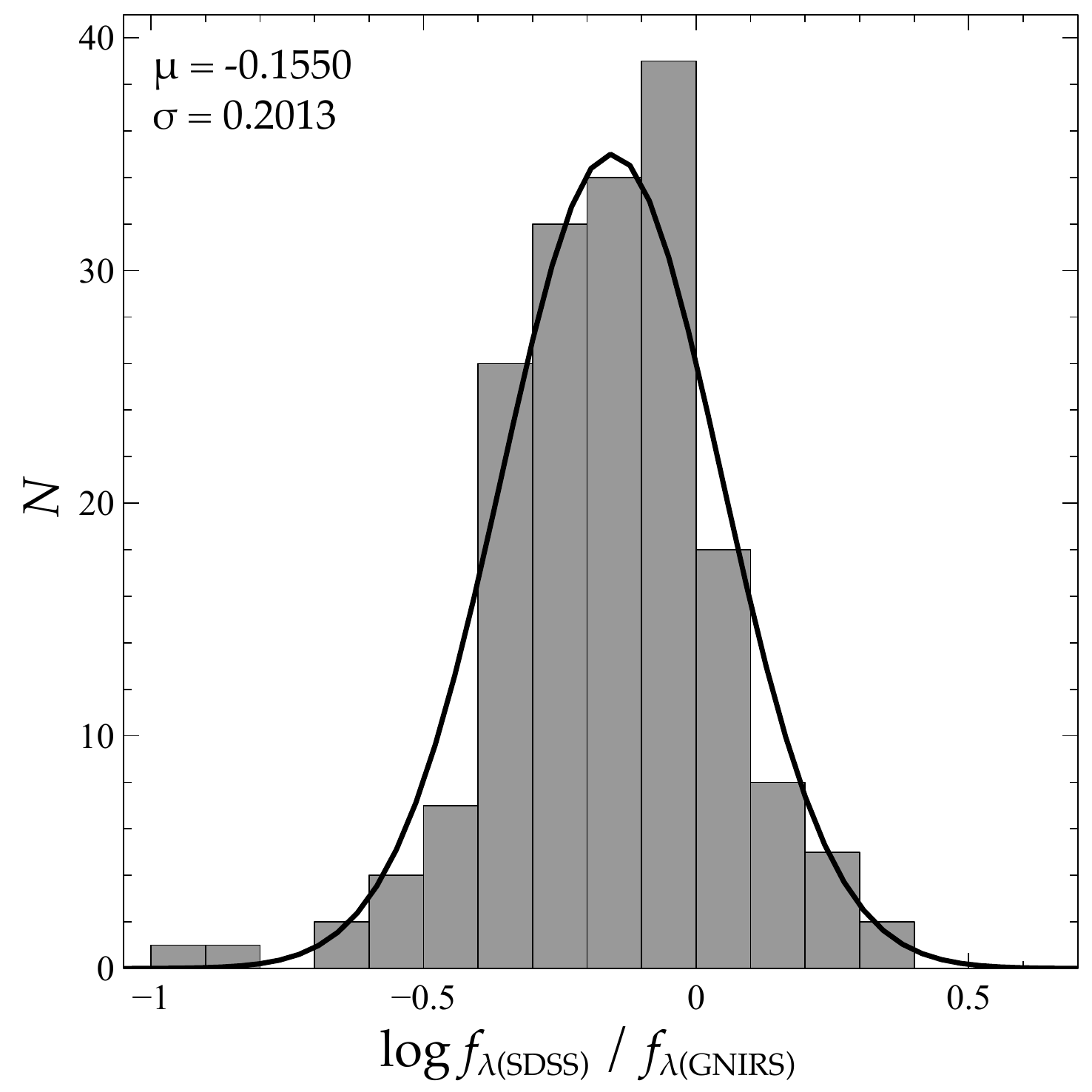}
\caption{Flux-density ratio distribution between SDSS and GNIRS spectra from the overlapping continuum regions ($\lambda_{\rm obs} \sim 8000 - 10000$ \AA) with a lognormal distribution fit. The log of the mean ratio ($\mu$) and its standard deviation ($\sigma$) indicate that the flux densities of the GNIRS spectra are consistent at the 1$\sigma$ level with those from their respective SDSS spectra.}
\label{fig:flux_hist}
\end{figure}

\begin{figure*}
\includegraphics[scale=0.6]{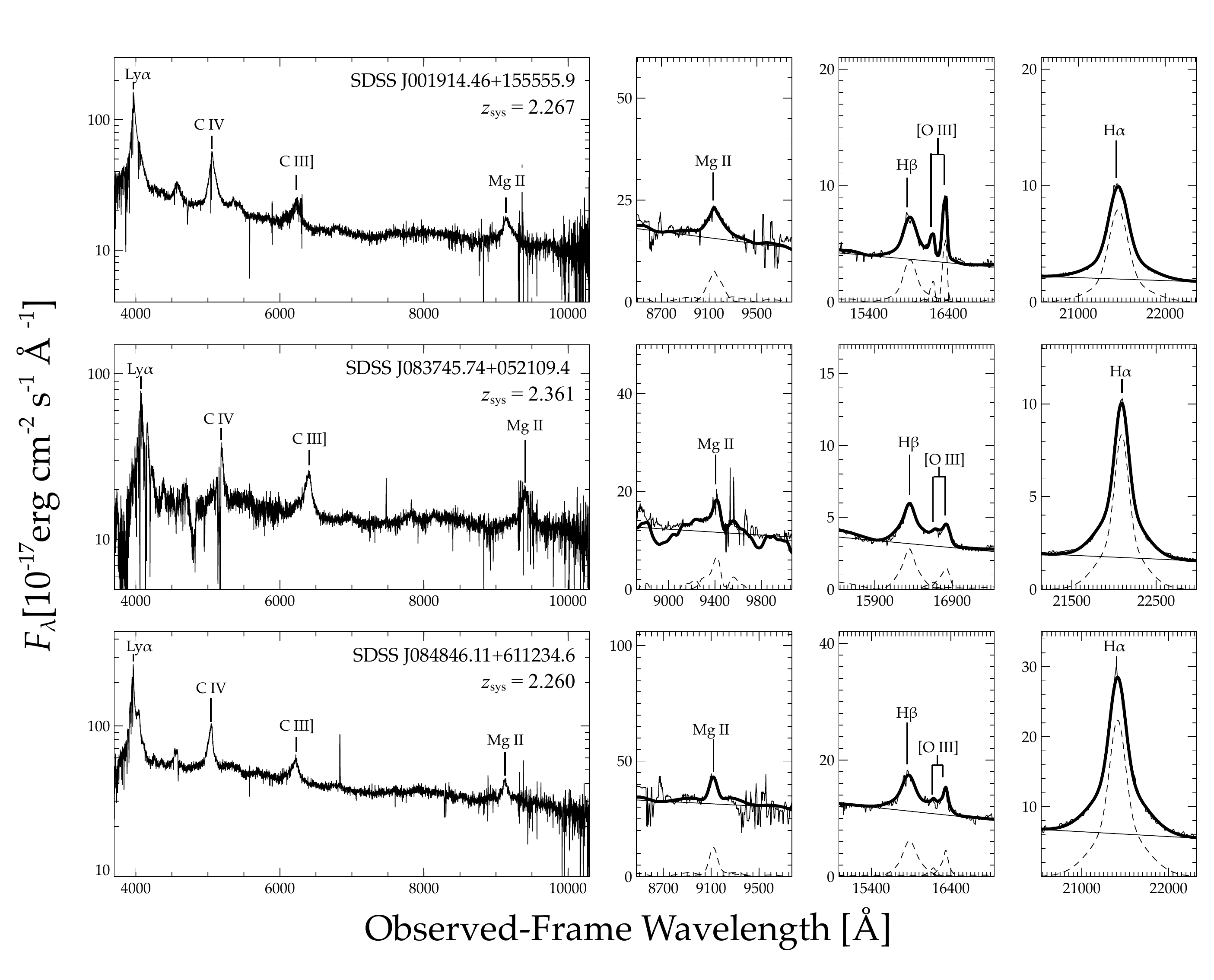}
\caption{SDSS and GNIRS spectra and their best-fit models for three representative quasars in our sample (fitting of the SDSS spectra is deferred to a future publication). From left to right, panels show the corresponding SDSS spectra, followed by the GNIRS \mgtwo, \hb, and \ha ~spectral regions, respectively. In the three rightmost panels, the spectrum is presented by a thin solid line, and best-fit models for the localized linear continua, Gaussian profiles, and iron emission blends are marked by dashed lines. Summed best-fit model spectra are overplotted with thick solid lines. Details of the spectral fitting procedure are given in Section~\ref{sec:spec}. All of the GNIRS spectra and their best-fit models are available electronically at \hbox{\url{http://physics.uwyo.edu/agn/GNIRS-DQS/spectra.html}}. We note that SDSS J083745.74+052109.4 is flagged as a BAL quasar (see, Table~\ref{tab:obslog}, \citeauthor{2017A&A...597A..79P} \citeyear{2017A&A...597A..79P}), and will be discussed in a future publication.}
\label{fig:fits}
\end{figure*}

In order to further test the reliability of our flux calibration, we compared the flux densities in overlapping continuum regions, $\lambda_{\rm obs} \sim 8000 - 10000$ \AA, between our GNIRS spectra and those of the respective SDSS spectra; this test was feasible for $\sim90\%$ of our sources that have both high-quality GNIRS and SDSS spectral data where we can obtain meaningful comparisons that avoid reductions in quality that can occur in this region for both surveys. We found that the flux densities in the SDSS spectra are, on average, smaller than the GNIRS flux densities by $\sim40\%$ ($\mu = -0.155$), with a 1$\sigma$ scatter of $\sim60$\% ($\sigma = 0.2013$) (see, Fig.~\ref{fig:flux_hist}, where $\mu$ and $\sigma$ are the logarithms of the mean and standard deviation, respectively). Therefore, the flux densities when directly comparing both spectral sets are consistent at the 1$\sigma$ level, despite the presence of this systematic offset. This systematic offset should be taken into account when comparing fluxes between SDSS and GNIRS spectra, however, it does not affect the emission-line measurements presented in this survey. This scatter may include discrepancies such as those due to intrinsic quasar variability, fiber light loss in SDSS spectra, and differences in airmass between quasars and their respective standard star observations. Examples of prominent emission lines in final, flux-calibrated spectra appear in Figure~\ref{fig:fits}.

\section{Spectral Fitting} \label{sec:spec}

The final GNIRS quasar spectra were fit by using multiple localized linear continua, explained in Section~\ref{sec:catalog}, constrained by no less than six narrow ($\sim200$\,\AA-wide, rest-frame) line-free regions, and performed Gaussian fits to the emission lines. The \fetwo ~and \fethree ~emission complexes were modeled via empirical templates from Boroson, \& Green (\citeyear[][]{1992ApJS...80..109B}) and Vestergaard \& Wilkes (\citeyear{2001ApJS..134....1V}) for the rest-frame optical and UV band, respectively. These templates were scaled and broadened by convolving a Gaussian with a full width at half maximum (FWHM) value that was free to vary between 1300 and 10000 km s$^{-1}$. Given that the \fetwo, \fethree, and \hb ~lines likely originate from different physical regions \citep[e.g.,][]{2013ApJ...769..128B}, we kept the FWHM of the iron templates as a free parameter. The FWHM values selected to broaden each template were determined using a least squares analysis on each fitted region. 

For the \othree ~lines, the widths of each line were restricted to be identical to each other, and their flux ratios were kept constant at $I_{5007} / I_{4959}  =  3$ \citep[e.g.,][and references therein]{2000MNRAS.312..813S}; additionally, the rest-frame wavelength difference between the $\lambda5007$ and $\lambda4959$ lines was kept constant, which proved adequate for the fits of each object. 

We fit two Gaussians to each broad emission line profile to accommodate possible asymmetry present in the profile due to, e.g., absorption, or outflows. We note that the two Gaussians fit per broad emission-line are adopted for fitting purposes only, and they do not represent physically distinct regions. Fitting the line profiles with more complex models was not warranted given the quality of our GNIRS spectra. The constraints on the Gaussian profiles for each emission line were that the peak wavelengths can differ from their known rest-frame values by up to \hbox{$\pm$ 1500 {\rm km s}$^{-1}$,} on initial assessment \citep[see, e.g.,][Figure 5]{2011AJ....141..167R} with a max flux value ranging from zero to a value calculated to be twice the maximum value of the emission line. Visual inspection yielded some exceptions beyond an offset of \hbox{$\pm$ 1500 {\rm km s}$^{-1}$}, whereupon manual fitting was performed to compensate for the larger velocity offset.

\subsection{Continuum Fitting} \label{sec:catalog}

By using localized linear continuum fitting, we were able to achieve more accurate measurements by avoiding uncertainties stemming from a single power-law fit. There has been debate about an accurate model for quasar continua: a single power-law, a broken power-law \citep[e.g.,][]{2001AJ....122..549V}, or whether the power-law description is appropriate at all in the rest-frame optical band; for example, in highly reddened quasar spectra a single power-law fitting will likely fail \citep[see, e.g.,][]{2019ApJS..241...34S}. Alternatively, quasar continua may be better described by accretion disk modeling \citep[e.g.,][]{2016MNRAS.460..187M}. This survey was primarily concerned with measuring emission-line properties as opposed to continua, and, through using a variety of fitting methods including our own investigations into the efficacy of power-law and broken power-law fits, we conclude that localized linear continua give, at worst, the same level of uncertainty as power-law fitting, and, at best, avoid large uncertainties inherent in modeling blended continuum features. Therefore, measurements of all the emission lines implemented localized linear continua where the windows for fitting were determined by the availability of the nearest continuum band segments as defined in \cite{2001AJ....122..549V}.

\subsection{\mgtwo} \label{sec:mg2}

The \mgtwo ~doublet is detected in the bluer regions of our spectra, where the $S/N$ is lower by roughly an order of magnitude than the redder regions where the \hb ~line is detected. Since our survey was designed such that the $S/N$ near the \hb ~region would be roughly comparable to the $S/N$ across the respective SDSS spectrum of each source (see, Section~\ref{sec:targets}), the $S/N$ around the \mgtwo ~region in our GNIRS spectra is roughly an order of magnitude lower than the corresponding values in the SDSS spectra. As a result, we were only able to obtain reliable \mgtwo ~and \hbox{\fetwo+\fethree} ~fits for $\sim$31\% of our sources (and we do not present measurements for \hbox{\fetwo+\fethree} ~due to their considerable uncertainties). In this work, we only present \mgtwo ~line measurements based on the GNIRS spectra of our sources; in a future publication, we will complement these data with \mgtwo ~line measurements based on the sample's SDSS spectra (for $\sim$87\% of our sample at \hbox{$z \lesssim 2.4$}).  On average, the uncertainties on the measured \mgtwo ~properties are roughly an order of magnitude larger than those of \hb. During the fitting process, we made a preliminary evaluation of the noise around the \hb ~and \mgtwo ~lines.  If the noise around \mgtwo ~was within a defined threshold ($S/N\sim10$) when compared to that of the \hb ~region ($S/N\sim40$, see Section~\ref{sec:observations}), the \mgtwo ~line was fit automatically. Otherwise, each spectrum was visually inspected to determine if it was possible to perform reliable measurements of the \mgtwo ~line. Due to the lower $S/N$ in this region, the \fetwo+\fethree ~complex was fit with narrow \hbox{($\sim20$~\AA)} continuum bands and often required further interactive adjustments in order to avoid noise spikes to ensure accurate fitting to the \mgtwo ~feature.

\subsection{\hb} \label{sec:hb}

The \hb ~region, for most of our objects, provided reliable measurements given the survey was designed with this region in mind. However, in $\lesssim$ 2\% of our objects, the \hb ~emission line was adjacent to the edge of the observing band, resulting in larger uncertainties when fitting the \fetwo ~emission complex. This misalignment of \hb ~stems from selecting our sample using UV-based redshifts, based primarily on the peak wavelength of the \cfour ~emission line, which suffer from systematic biases due to outflows that can be as large as \hbox{$\approx 5000$ $\rm{km}$ $\rm{s^{-1}}$} \citep[][Matthews et al., in prep., 2021]{2020ApJ...893...14D}. This misalignment also results in reduced coverage of the \fetwo ~blends for these objects. Despite this complication, we were able to adequately fit two Gaussians to each of the \hb ~emission lines.

\begin{figure}[H]
\includegraphics[scale=0.65]{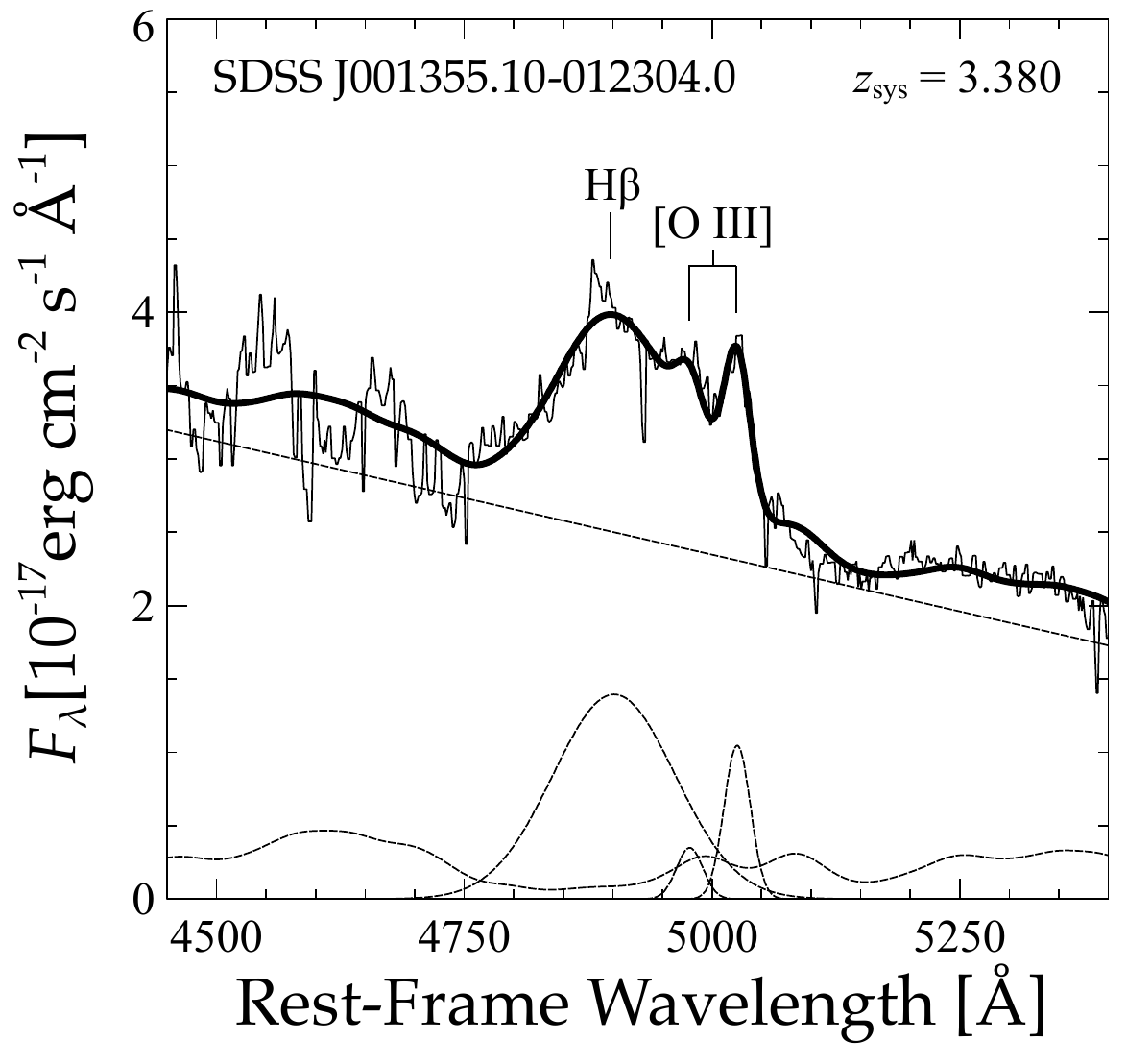}
\caption{GNIRS spectrum of the \hb ~region of \hbox{SDSS J001315.10-012304.0}, $z_{sys} = 3.380$. The ``shelf'' structure redward of the \hb ~line appears to be a result of strong \fetwo ~and mild \othree ~emission. This differs from typical ``Eigenvector 1'' trends in Boroson, \& Green (\citeyear{1992ApJS...80..109B}), where sources with strong \fetwo ~blends tend to have weak \othree ~lines. Line styles are as in Fig.~\ref{fig:fits}. These shelves may be a signature of binary quasar candidates \citep[see, e.g.,][]{1994ApJS...90....1E}}
\label{fig:shelf}
\end{figure}

By design, our survey targeted highly luminous quasars, biased toward having higher $L/L_{\rm Edd}$ values \citep[see, Fig.~\ref{fig:sdss},][]{2003ApJ...583L...5N}, which typically also tend to have relatively strong \fetwo ~emission. As a result, we relied on the broad iron bumps on either side of the \hb ~line, rest-frame $\sim  4450 - 4750$~\AA ~and $5100 - 5400$~\AA\ \citep{2001AJ....122..549V}, as our primary region for fitting the \fetwo ~complex. While reasonable in most cases, these fits are likely affected by \hetwo ~$\lambda 4686$ emission-line contamination, however the \hetwo ~emission line is unresolvable in this sample due to uncertainties from a variety of factors (see Section~\ref{sec:err}). On average, the corresponding \fetwo ~EW values in those sources is \hbox{$\sim20$~\AA}. Additionally, \hbox{$\sim$ 5\%} of our objects differed from the well-known trends of ``Eigenvector 1" \citep{1992ApJS...80..109B}, having a blend of strong \othree ~and \fetwo ~emission, resulting in their spectra exhibiting ``shelves'' on the red side of the \hb ~profile. These features required a more careful fitting, and we did not see any evidence of \othree ~outflows directly contributing to this emission complex.  An example of a shelf-like fit is presented in Figure~\ref{fig:shelf}. 
 
\begin{figure}[H]
\includegraphics[scale=0.5]{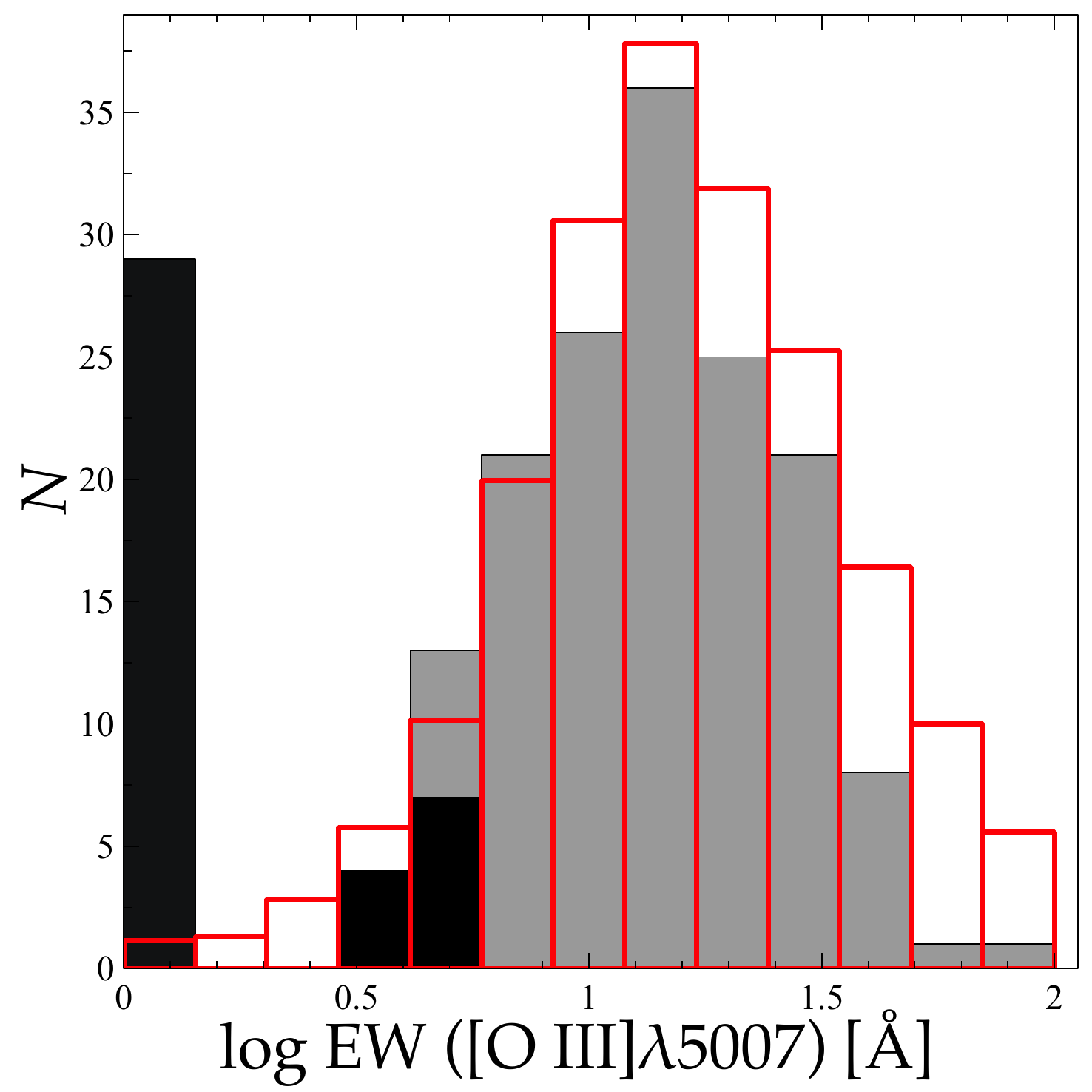}
\caption{\othree ~rest-frame EW distribution of the GNIRS-DQS sources (grey) overplotted with rest-frame \othree ~EWs from \cite{2011ApJS..194...45S} (red outline; scaled down by a factor of $100$). For $\sim19$\% of the GNIRS-DQS sources that lack detectable \othree ~emission we are able to place strong upper limits on their EW values (black). When compared to \othree ~measurements of low-redshift, low-luminosity sources from \cite{2011ApJS..194...45S}, the \othree ~emission tends to become weaker as luminosity increases, consistent with the trends observed in previous studies of high-redshift quasars \citep[e.g.,][]{2004ApJ...614..558N,2016ApJ...817...55S}.}
\label{fig:O3EW}
\end{figure}

Figure~\ref{fig:O3EW} shows the distribution of \othree ~EWs in the GNIRS-DQS sample. As explained in Section~\ref{sec:the} below, for those objects that do not have detectable \othree ~emission, we must use the \mgtwo ~line to determine systemic redshifts ($z_{\rm sys}$); for those objects that lack both \othree ~and \mgtwo, we must utilize the \hb ~line for that purpose, which is present in every GNIRS-DQS spectrum.

\subsection{\ha}

Being the most prominent feature in all the spectra of our sources at \hbox{$z < 2.5$} (constituting $\sim87\%$ of the sample), \ha ~yielded the smallest uncertainties on all the emission-line parameters.  We do not detect significant narrow \ntwo ~emission-lines flanking the \ha ~line in any of our sources, which is expected given our selection of highly luminous quasars \citep[e.g.,][]{1986ApJ...302...56W,2011ApJS..194...45S}.

\subsection{Uncertainties in Spectral Measurements} \label{sec:err}

The uncertainties inherent in the GNIRS spectra are contributed by a variety of factors. These include (but are not limited to) sub-par observing conditions, the use of replacement flat fields in several of the spectra (see Section~\ref{sec:observations}), and differences in airmass and/or atmospheric conditions between the standard star and the respective quasar observations. Moreover, modeling the telluric standard star continuum with a blackbody function fails to account for potential NIR excess emission from a circumstellar disk around the star. These factors lead to uncertainties on the flux density and shapes of the emission-line profiles, including the locations of their peaks. The uncertainties on these parameters are in the range \hbox{$\approx$ 4-7$\%$}, \hbox{$\approx$ 3-6$\%$}, \hbox{$\approx$ 2-5$\%$}, and \hbox{$\approx$ 2-4$\%$}, for each emission line, respectively. On average, these uncertainties result in general measurement errors across all parameters for an emission-line profile of up to \hbox{$\sim$ 7\%}.

Emission-line fitting first relied on shifting the spectrum to the rest-frame using the best available SDSS redshift.  However, due to inaccuracies with the SDSS redshift, the emission-lines in the GNIRS spectra often did not line up with the known rest-frame values.  This offset led to uncertainties during fitting, and was ultimately mitigated by introducing a redshift iteration process.  Emission-lines were fit for three different regimes separately, the \mgtwo, \hb, and \ha ~regions, based off of the SDSS redshift. A systemic redshift, $z_{sys}$, was then determined by the best fit of the most reliable emission-line for measuring redshift, as discussed in Section~\ref{sec:the} below, and the spectrum was shifted according to this value. This process was repeated until the difference in consecutive redshifts was less than \hbox{$z_{n-1} - z_n < 0.001$} for each region. Additionally, this redshift iteration allows more accurate measurements on $z_{sys}$, the flux density at rest-frame \hbox{5100 \AA} ~($F_{\lambda,5100}$), and more accurate fitting of the broadened iron templates.

After identifying the most accurate redshift, final fits are performed on emission-line features. Using preliminary Gaussian and localized linear continuum fits, residuals are generated, which yield upper and lower values for uncertainties present across the fitting region. With these residual bounds, Gaussian noise is introduced, and a series of 50 fits is performed in order to generate upper and lower bound estimates on the final Gaussian fits. To quantify the error on best-fit parameters, each iterated fit value is stored, which is used to generate a distribution of principle measurements. These distributions are then fit using a Gaussian function in order to determine the final errors at a 1$\sigma$ confidence level. The iron templates of the \hb ~and \mgtwo ~regions also experience iterations of FWHM for the line profile, which allows for accurate \fetwo ~and \fethree ~broadening error estimates. These various fitting iterations allow conservative error estimates on basic emission-line parameters, i.e., FWHM, EW, and line peaks. Finally, the best fit spectral model for each source was verified by visual inspection. 

\subsection{The Catalog} \label{sec:the}

Table~\ref{tab:meas} describes the format of the data presented in the catalog. It contains basic emission line properties, particularly the FWHM and rest-frame EW, of the \mgtwo, \hb, \othree, and \ha ~emission lines. The catalog also provides observed-frame wavelengths of emission-line peaks, as well as the asymmetry and kurtosis of each emission line, which were obtained from the Gaussian fits. A host of additional parameters are given, including the FWHM of the kernel Gaussian used for broadening the \fetwo\ blends around the \hb ~region and the EW of these blends in the $4434 - 4684$ \AA ~region (following \citeauthor{1992ApJS...80..109B} \citeyear{1992ApJS...80..109B}), as well as the flux density and monochromatic luminosity ($\lambda L_{\lambda}$) at 5100 \AA. The catalog also provides $z_{\rm sys}$ values measured from observed-frame wavelengths of emission-line peaks. For determining $z_{\rm sys}$, we adopt the observed-frame wavelength of the peak of one of three emission lines with the highest degree of accuracy which is present in the GNIRS spectrum, where it is known that these three emission lines have uncertainties of \hbox{~$\simeq$ $50$ $\rm{km}$ $\rm{s^{-1}}$}, \hbox{~$\simeq$ $200$ $\rm{km}$ $\rm{s^{-1}}$}, and \hbox{~$\simeq$ $400$ $\rm{km}$ $\rm{s^{-1}}$} for \othree, \mgtwo, and \hb, respectively (\citeauthor*[][]{2016ApJ...831....7S} \citeyear[][]{2016ApJ...831....7S}).

In cases where the prominent emission lines (i.e., \mgtwo, \hb, \othree, ~and \ha) have no significant detections, upper limits are placed on their EWs by assuming FWHM values for each line using the median value in the sample distributions, and taking the weakest feature detectable in the GNIRS spectra for each line. Additionally, we placed upper limits on the EW of the optical \fetwo\ blends in cases where excess noise surrounding the \hb+\othree ~region would not enable us to fit the \fetwo\ blends reliably; we found that a value of 2 \AA\ for this parameter provides a conservative upper limit in all such cases.

Finally, additional, and typically weaker, emission line measurements follow the formatting presented in Table~\ref{tab:suppmeas}, and are reported in the supplemental features catalog for 106 sources from our sample where such features could be measured reliably. These emission lines were fit on a case-by-case basis after visually inspecting each GNIRS spectrum (and no upper limits are assigned in cases of non-detections). Where applicable, we performed fits on the following emission lines with two Gaussians per line, following the same methodology used for primary emission line measurements: \hd~$\lambda 4101$, \hg~$\lambda 4340$, and \nethree~$\lambda 3871$. The \otwo~$\lambda 3727$ doublet was fit in the same manner.

\section{Summary} \label{sec:app}

We present a catalog of spectroscopic properties obtained from NIR observations of a uniform, flux-limited sample of 226 SDSS quasars at \hbox{$1.5 \lesssim z \lesssim 3.5$}, which is the largest, uniform inventory for such sources to date. The catalog includes basic spectral properties of \mgtwo, \hb, \othree, \fetwo, and \ha ~emission lines, as well as \hd, \hg, \otwo, and \nethree ~emission lines for a subset of the sample. A spectral resolution of $R\sim1,100$ was achieved for this data set, which is roughly comparable to the value of the corresponding SDSS spectra. These measurements provide a database to comprehensively analyze and investigate rest-frame UV-optical spectral properties for high-redshift, high-luminosity quasars in a manner consistent with studies of low-redshift quasars.

In particular, the catalog will enable future work on robust calibrations of UV-based proxies to systemic redshifts and black-hole masses in distant quasars. Such prescriptions are becoming increasingly more important as millions of quasar optical spectra will be obtained in the near future by, e.g., the Dark Energy Spectroscopic Instrument \citep[DESI;][]{2013arXiv1308.0847L,2016arXiv161100036D} and the 4-metre Multi-Object Spectroscopic Telescope \citep[4MOST;][]{2012SPIE.8446E..0TD}, where reliable estimates of $z_{sys}$ and $M_{\rm BH}$ will be crucial to extract the science value from these surveys. In forthcoming papers we will present, among other facets, redshift calibrations via indicative emission lines such as \othree ~(Matthews et al., in prep.), SMBH estimates using the \hb ~and \mgtwo ~profiles measured in this survey (Dix et al., in prep.), and correlations among UV-optical emission lines \citep[e.g.,][]{1992ApJS...80..109B,1999ApJ...515L..53W,2012ApJ...753..125S}.

In the future, we should continue to push the redshift barrier for the \hb ~and \othree ~emission lines, as current investigations have been confined to $z\lesssim3.5$, in order to gain an increased understanding of the co-evolution of SMBHs and their host galaxies, along with more reliable redshifts. However, at redshifts higher than $z\sim3.5$, these observations cannot be obtained via ground-based telescopes. Future studies in this respect could include a two-pronged approach using small calibration surveys. The first survey, for example, can use higher resolution instruments such as Gemini's Spectrograph and Camera for Observations of Rapid Phenomena in the Infrared and Optical \citep[SCORPIO;][]{2018SPIE10702E..0IR} which will better measure weak emission-line profiles and obtain more accurate measurements of the prominent emission lines. This information will reinforce the measurements of this survey and allow for more confident applications to much higher redshifts, even beyond $z > 6$. The second survey would be a select sample of a few dozen highly luminous $z > 3.5$ objects using space-based observations from the James Webb Space Telescope \citep[JWST,][]{2006SSRv..123..485G} for optimal spectral quality, with the possibility for a contemporaneous SCORPIO survey to obtain measurements of lines such as \cfour ~from the ground.

\acknowledgments

This work is supported by National Science Foundation grants AST-1815281 (B.~M.~M., O.~S., C.~D.), AST-1815645 (M.~S.~B., A.~D.~M.), AST-1516784 (W.~N.~B.), and AST-1715579 (Y.~S.). A.D.M. was also supported, in part, by the Director, Office of Science, Office of High Energy Physics of the U.S. Department of Energy under Award No. DE-SC0019022. Y.~S. acknowledges support from an Alfred P. Sloan Research Fellowship. We thank an anonymous referee for thoughtful and valuable comments that helped improve this manuscript. This research has made use of the NASA/IPAC Extragalactic Database (NED), which is operated by the Jet Propulsion Laboratory, California Institute of Technology, under contract with the National Aeronautics and Space Administration. We thank Jin Wu and the Gemini Observatory staff for helpful discussions, and providing assistance with data reduction for this project. This work was enabled by observations made from the Gemini North telescope, located within the Maunakea Science Reserve and adjacent to the summit of Maunakea. We are grateful for the privilege of observing the Universe from a place that is unique in both its astronomical quality and its cultural significance.

\startlongtable
\begin{deluxetable*}{lcccccccccc}
\tablecolumns{9}
\tabletypesize{\scriptsize}
\tablewidth{0pc}
\tablecaption{Observation Log \label{tab:obslog}} 
\tablehead
{
\colhead{Quasar} &
\colhead{$z_{\rm SDSS}$\tablenotemark{a}} &
\colhead{$J$} &
\colhead{$H$} &
\colhead{$K$} &
\colhead{Obs. Date} &
\colhead{Semester} &
\colhead{Net Exp.} &
\colhead{Comments} &
\colhead{BAL} &
\colhead{RL} \\
\colhead{} &
\colhead{} &
\colhead{[mag]} &
\colhead{[mag]} &
\colhead{[mag]} &
\colhead{} &
\colhead{} &
\colhead{[s]} &
\colhead{} &
\colhead{} &
\colhead{} \\
\colhead{(1)} &
\colhead{(2)} &
\colhead{(3)} &
\colhead{(4)} &
\colhead{(5)} &
\colhead{(6)} &
\colhead{(7)} &
\colhead{(8)} &
\colhead{(9)} &
\colhead{(10)} &
\colhead{(11)} 
}
\startdata
SDSS J000544.71$-$044915.2  & 2.322 & 16.94  & 16.09  & 16.66  &     2019 Oct 18                &   2019B    &   1800       &		4   & \nodata & \nodata   \\
SDSS J000730.94$-$095831.5 & 2.223   & 17.09  & 15.94  & 15.37  &  2019 Jan 06               & 2018B    & 1800       &		4  & 1 & \nodata   \\
SDSS J001249.89+285552.6 & 3.236    & 16.51   & 15.71   & 15.49   &  2017 Sep 09                & 2017B    & 1800       &		\nodata      & 1 & \nodata      \\
SDSS J001355.10$-$012304.0 & 3.396     & 16.71  & 16.05  & 15.46  &  2019 Jan 05 & 2018B    & 900                  &		\nodata   & 1 & \nodata  \\
&\nodata & \nodata      & \nodata  & \nodata  & 2019 Jan 07 & 2018B    & 900        &		\nodata     & \nodata & \nodata          \\
SDSS J001453.20+091217.6 & 2.338    & 16.65  & 15.92  & 15.14  &  2017 Sep 19               & 2017B    & 2025       &		1    & \nodata & \nodata     \\
SDSS J001813.30+361058.6 & 2.316       & 16.15  & 15.65  & 14.75  & 2017 Aug 31               & 2017B    & 1800        &		\nodata     & \nodata & \nodata     \\
SDSS J001914.46+155555.9 & 2.271    & 16.72  & 15.81  & 15.14  & 2017 Sep 01                & 2017B    & 1800       &		\nodata        & \nodata & \nodata     \\
\enddata
\tablenotetext{a}{Value based on best available measurement as stated by SDSS (P\^aris et al. 2018, Table A1, column 9 ``Z").}
\tablecomments{Objects followed by an empty row aside from observation date, semester, and net exposure are additional observations made for that same object.\\
Comments:\\
(1) At least one exposure was taken under subpar observing conditions.\\
(2) All exposures were taken under supbar observing conditions.\\
(3) Supplemental data used from other observations to aid in reduction as described in Section~\ref{sec:err}.\\
(4) Observation failed to provide spectrum of the source due to bad weather, instrument artifacts, or other technical difficulties during the observation.}
\end{deluxetable*}

\begin{deluxetable*}{llllll}
\tablecolumns{6}
\tabletypesize{\scriptsize}
\tablewidth{0pc}
\tablecaption{Column Headings for Spectral Measurements \label{tab:meas}} 
\tablehead
{
\colhead{Column} &
\colhead{Name} &
\colhead{Bytes} &
\colhead{Format} &
\colhead{Units} &
\colhead{Description}\\
\colhead{(1)} &
\colhead{(2)} &
\colhead{(3)} &
\colhead{(4)} &
\colhead{(5)} &
\colhead{(6)} 
}
\startdata
1 & OBJ & (1-24) & A24 & \nodata & SDSS object designation\\
2 & ZSYS & (26-29) & F4.3 & \nodata & Systemic redshifts\\
3 & LC\_MG II & (31-35) & F5.0 & \AA & Observed-frame wavelength of the emission line peak of \mgtwo ~based on peak fit value\\
4 & LC\_MG II\_UPP & (37-39) & F3.0 & \AA & Upper uncertainty for the line peak of \mgtwo\\
5 & LC\_MG II\_LOW & (41-43) & F3.0 & \AA & Lower uncertainty for the line peak of \mgtwo\\
6 & FWHM\_MG II & (45-48) & F4.0 & $\rm{km}$ $\rm{s^{-1}}$  & FWHM of \mgtwo\\
7 & FWHM\_MG II\_UPP & (50-52) & F3.0 & $\rm{km}$ $\rm{s^{-1}}$  & Upper uncertainty of FWHM of \mgtwo\\
8 & FWHM\_MG II\_LOW & (54-56) & F3.0 & $\rm{km}$ $\rm{s^{-1}}$  & Lower uncertainty of FWHM of \mgtwo\\
\enddata
\tablecomments{Data formatting used for the catalog. Asymmetry is defined here as the skewness of the Gaussian fits, i.e., a measure of the asymmetry of the distribution about its mean, $s = E(x - \mu)^3 / \sigma^3$, where $\mu$ ~is the mean of $x$, $\sigma$ ~is the standard deviation of $x$, and $E(t)$ is the expectation value. Kurtosis is the quantification of the "tails" of the Gaussian fits defined as $k = E(x - \mu)^4 / \sigma^4$, where symbols are the same as for asymmetry.}
\tablenotetext{a}{A value of 2~\AA\ denotes an upper limit on this parameter.}
\end{deluxetable*}

\begin{deluxetable*}{llllll}
\tablecolumns{6}
\tabletypesize{\scriptsize}
\tablewidth{0pc}
\tablecaption{Column Headings for Supplemental Emission-Line Measurements \label{tab:suppmeas}} 
\tablehead
{
\colhead{Column} &
\colhead{Name} &
\colhead{Bytes} &
\colhead{Format} &
\colhead{Units} &
\colhead{Description}\\
\colhead{(1)} &
\colhead{(2)} &
\colhead{(3)} &
\colhead{(4)} &
\colhead{(5)} &
\colhead{(6)}
}
\startdata
1 & OBJ & (1-24) & A24 & \nodata & SDSS object designation\\
2 & LC\_HD & (26-30) & F5.0 & \AA & Observed-frame wavelength of the emission line peak of \hd ~based on peak fit value\\
3 & LC\_HD\_UPP & (32-35) & F4.0 & \AA & Upper uncertainty for the line peak of \hd\\
4 & LC\_HD\_LOW & (37-40) & F4.0 & \AA & Lower uncertainty for the line peak of \hd\\
5 & FWHM\_HD & (42-45) & F4.0 & $\rm{km}$ $\rm{s^{-1}}$  & FWHM of \hd\\
6 & FWHM\_HD\_UPP & (47-49) & F3.0 & $\rm{km}$ $\rm{s^{-1}}$  & Upper uncertainty of FWHM of \hd\\
7 & FWHM\_HD\_LOW & (51-53) & F3.0 & $\rm{km}$ $\rm{s^{-1}}$  & Lower uncertainty of FWHM of \hd\\
8 & EW\_HD & (55-56) & F2.0 & \AA & Rest-frame EW of \hd\\
\enddata
\tablecomments{Data formatting used for the supplemental measurements in the supplemental features catalog.}
\tablenotetext{a}{\otwo ~$\lambda$ 3727}
\tablenotetext{b}{\nethree ~$\lambda$ 3870}
\end{deluxetable*}

\end{document}